\documentstyle[12pt]{article}
\newcommand{\beq}{\begin{equation}}
\newcommand{\eeq}{\end{equation}}
\newcommand{\bra}{\begin{array}}
\newcommand{\era}{\end{array}}

\begin{document}
\begin{center}
{\Large\bf Between Quantum Virasoro Algebra ${\mathcal{L}}_{c}$
and Generalized Clifford Algebras}
 \vskip1.4truecm {\bf E. H. El
Kinani}\footnote{ Junior Associate at The Abdus Salam ICTP,
Trieste, Italy. \\Permanent address, Universit\'e Moulay Ismail,
Facult\'e des Sciences et Techniques, D\'epartement de
Math\'ematiques, Groupe de Physique
Math\'ematique, Boutalamine B.P.509, Errachidia, Morocco.}\\
 The Abdus Salam International Center for Theoretical Physics, Strada Costiera 11, 34014, Trieste, Italy \\
\end{center}

\begin{abstract}
\hspace{.3in}In this paper we construct the quantum Virasoro
algebra ${\mathcal{L}}_{c}$ generators in terms of operators of
the generalized Clifford algebras $C_{n}^{k}$. Precisely, we show
that ${\mathcal{L}}_{c}$ can be embedded into generalized Clifford
algebras.
\end{abstract}
\newpage

\section{Introduction}
\hspace{.3in} Over the past decade, much attention has been paid
to the generalized Clifford algebras and its associated Grassmann
algebras in connection with the important role investigated in
mathematics and physics (see e.g \cite{1,2}). In fact, the
linearization of the homogeneous polynomials of degree $n$ and $k$
variables leads to generalized Clifford algebras $C_{n}^{k}$ and
to the introduction of the multicomplex plane ${\mathcal{MC}}_{n}$
where the fundamental element $e$ satisfies the basic relation
$e^{n}=-1$ \cite{3}. On the other hand, the generalized Grassmann
algebras which have been introduced firstly in framework of the 2D
conformal field theories and they arise also in the contexts of
quantum groups and fractional generalization of supersymmetry
quantum mechanics \cite{4} and next they were used for
generalization of supersynmmetry to the fractional supersynmmetry
\cite{5} which have found applications to various physical
problems, such as the
description of the fractional statistics (see e.g \cite{6})\\

In this letter, we purpose the construction of the quantum
conformal symmetry ${\mathcal{L}}_{c}$ in terms of the generalized
Clifford generators $\gamma_{i}$. In fact, in a previous paper
\cite{7}, we have done the embedding  of the classical Virasoro
algebra ${\mathcal{L}}_{0}$ (with out central charge $c$) into
generalized Clifford algebras here we will extend this embedding
to the case of quantum Virasoro algebra ${\mathcal{L}}_{c}$ (with
central charge $c$). First, we will recall some notions connected
with the generalized Clifford algebras and its associated
generalized Grassmann algebras. Then, after recalling the
classical and quantum conformal symmetries, we will present our
embedding of ${\mathcal{L}}_{c}$ algebra into $C_{n}^{k}$. In
fact, we will show that the generators of ${\mathcal{L}}_{c}$ may
be expressed via the generators $\gamma_{i}$ of the generalized
Clifford algebras.

\section {Preliminaries on the generalized Clifford Algebras}
In brief, the $n$-dimensional generalized Clifford algebras of
order $k$ (GCA) is defined as an associative algebra over complex
numbers $\mathcal{C}$, generated by operators (matrices)
$\gamma_{i}$
 (for more details see \cite{1, 2} )satisfying the following
 commutation relations \beq\gamma_{i}\gamma_{j}=\omega\gamma_{j}
 \gamma_{i}, \:\:\:\:\ i<j\:\:\ (i,j =1,2,...n)\eeq and \beq
 (\gamma_{i})^{k}=1 \:\:\:\:\
i=1,2,...n \eeq where $\omega=e^{\frac{2\pi i}{k}}$ is an $k$th
primitive root of unity.\\

The generators $\{\gamma_{i} \}^{n}_{1}$ can be represented as
tensor product of the generalized Pauli matrices  \cite{1,2} in
complete analogy with the usual Clifford algebra. If we substitute
the equation (2) by $(\gamma_{i})^{k}=0$, we obtain the so-called
generalized Grassmann algebras which have been introduced firstly
in framework of the 2D conformal field theories and were
rediscovered in the contexts of quantum groups and fractional
generalization of susy quantum mechanics and next these algebras
were used also for generalization of supersynmmetry to the
fractional supersynmmetry.

\section {Classical and Quantum Virasoro symmetries }
\hspace{.3in}For reader convenience we remind some information
about classical and quantum Virasoro symmetries termed also
conformal symmetries these latter were first introduced in the
context of string theories and they are relevant to any theory in
(2+1)-dimensional space-time which possesses conformal invariance.
The conformal symmetry consists of all general transformations
\begin{equation}
z\to z+\epsilon(z), \:\:\:\:\ \bar{z} \to \bar{z} +
\bar{\epsilon}(\bar{z})
\end{equation}
where $\epsilon(z)$ and $(\bar{\epsilon}(\bar{z}))$ are an
infinitesimal analytical (anti-analytical) functions. It can be
represented as an infinite Laurent series
\begin{equation}
\epsilon(z)=\sum_{n}\epsilon_{n}z^{n+1},  \:\:\:\:\ n \in
\mathcal{Z},
\end{equation}
and an analogous formula holds for $\bar{\epsilon}(\bar{z})$.\\

These mapping are generated by the differential operators
\begin{equation}
V_{m} = -z^{m+1}\partial_z  \:\:\  \mbox {and} \:\:\ \bar{V}_{m} =
- \bar{z}^{m+1}\partial_{\bar{z}}.
\end{equation}
The operators $\ell_{m}$ satisfy the commutator relations
\begin{equation}
[\ V_{m}, V_n] = (m-n)V_{m+n},
\end{equation}
and an analogous formula holds for the $\bar{V}_{m}$. Hence, this
means that both the $V_{m}$ and the $\bar{V}_{m}$ span an
infinite-dimensional Lie algebra. Moreover, these two algebras are
combined as a direct sum, $[V_{m},\bar{V}_{m}]=0$. The algebra
defined by (6) is known as the classical conformal or classical
Virasoro algebra. We shall denote that the quantum version of the
above algebra is obtained by adding the central terms which is
connected with the quantum theory. Then, it well-known that the
classical Virasoro admits a unique $1$-dimentional central
extension
\begin{equation}
{\mathcal{L}}_{c}= {\mathcal{L}} \oplus {\mathcal{C}}c,
\end{equation}
with the following commutation relations
\begin{equation}
[\ V_{m}, c] = 0, \:\:\:\:\:\  [V_{m}, V_n] = (m-n)V_{m+n} +c
{\frac{(m^{3}-m)}{12}}\delta_{m+n,0}
\end{equation}
where the value of the central charge $c$ is the parameter of the
theory in the quantum field theory context (e.g $c=1$ for the free
boson). The unitary representation of the above algebra are
well-know (see e.g \cite{8}).
\section {${\mathcal{L}}_{c}$ as a subalgebra of Generalized Clifford Algebras }
\hspace{.3in}First, recall that in a previous paper \cite{7}, we
have realized the centerless ${\mathcal{L}}_{0}$ Virasoro algebra
( alias classical Virasoro algebra)as subalgebra of the
generalized Clifford algebras as a linear combination of the
generators $\gamma_{i}$. In fact for any pair $(i,j)$ such that
$i<j$ we have : \beq V^{(i,j)}_{N}=\sum_{ \bar{m} \ne(k,k), \ne
(0,0)}D^{\bar{m}}_{N}\gamma^{m_{1}}_{i}\gamma^{m_{2}}_{j}, \eeq
where $\bar{m}=(m_{1},m_{2})$, and  generators
$\gamma^{m_{1}}_{i}\gamma^{m_{2}}_{j} \sim e^{\bar{m}}_{(i,j)}$
satisfy the so-called centerless area preserving algebras (for $k$
large and after a renormalization of generators )

\beq [e^{\bar{m}}_{(i,j)}, e^{\bar{n}}_{(i,j)}]=(\bar{m} \times
\bar{n})e^{\bar{m}+\bar{n}}_{(i,j)}, \eeq where $\bar{m} \times
\bar{n} = m_{1}n_{2}-m_{2}n_{1}$. Imposing that the generators
$V^{(i,j)}_{N}$ satisfying \beq [V^{(i,j)}_{N}, V^{(i,j)}_{N'}]=(
N-N')V^{(i,j)}_{N+N'}. \eeq Then, this is equivalent to the
following identities among the constants $D^{\bar{m}}_{N}$ namely
\beq \sum_{ \bar{m} \ne(k,k), \ne
(0,0)}D^{\bar{l}-\bar{m}}_{N'}D^{\bar{m}}_{N}(\bar{m} \times
\bar{l})=(N-N')D^{\bar{l}}_{N+N'} \:\:\:\ (\bar{l} \ne (0,0)).\eeq
The particular simple solution is chosen to be of the forme \beq
D^{\bar{m}}_{N}=\frac{(-1)^{m_{1}}}{m_{1}}\delta_{m_{2},N}
\:\:\:\:\ m_{1}\ne 0 \eeq \beq D^{\bar{m}}_{N}=0
:\:\:\:\:\:\:\:\:\:\:\:\:\:\:\:\:\:\:\:\:\:\:\:\:\ m_{1}=0. \eeq
In what follows, we will extend the work done below on
${\mathcal{L}}_{c}$ Virasoro algebra ( alias quantum Virasoro
algebra eq(8)). In deed, if we now impose that the generators
$e^{\bar{m}}_{(i,j)}$ satisfy the area preserving with cental
non-trivial charge \beq [e^{\bar{m}}_{(i,j)},
e^{\bar{n}}_{(i,j)}]=(\bar{m} \times
\bar{n})e^{\bar{m}+\bar{n}}_{(i,j)}+\bar{a}.
\bar{m}\delta_{\bar{m},\bar{n}}, \eeq where $\bar{a}.
\bar{m}=a_{1}m_{1}+a_{2}m_{2}$. Then,  the linear combination
eq(9) satisfy a quantum Virasoro ${\mathcal{L}}_{c}$ algebra
relation eq(8), in fact  \beq [V^{(i,j)}_{N}, V^{(i,j)}_{N'}]=(
N-N')V^{(i,j)}_{N+N'}+ D_{N,N'} \eeq with $D_{N,N'}$ is given by
\beq D_{N,N'}=\sum_{ \bar{m}\ne
(0,0)}D^{\bar{m}}_{N}D^{\bar{-m}}_{N'}\bar{a}. \bar{m}\eeq From
the Jacobi identity for the $e^{\bar{n}}_{(i,j)}$'s \beq
[e^{\bar{m}}_{(i,j)},[e^{\bar{n}}_{(i,j)},e^{\bar{l}}_{(i,j)}]]+\mbox{cyclic
perm.}=0\eeq the constant $D_{N,N'}$ are consistent with the
Jacobi identity for $V^{(i,j)}_{N}$ \beq
[V^{(i,j)}_{M},[V^{(i,j)}_{N},V^{(i,j)}_{L}]]+ \mbox{cyclic
perm.}=0\eeq Which imply that  \beq
(M-N)D_{L,N+M}+(M-L)D_{M,N+L}+(L-M)D_{N,L+M}=0\eeq and,
additionally, we have also the skew-symmetric relation \beq
D_{M,N}=-D_{N,M}\eeq Following \cite{9}, using the transformation
$V^{(i,j)}_{M} \to V^{(i,j)}_{M}+MV^{(i,j)}_{M}D_{M,0}$, where $M
\ne 0$ and $ V^{(i,j)}_{0} \to V^{(i,j)}_{0}+\frac{1}{2}D_{1,-1}$,
we obtain after wards $D_{M,0}=D_{0,M}=D_{1,-1}=0$. Then putting
$L=0$, we deduce that $D_{M,N}=0$ unless $M=-N$ and putting
$L=-M-1$ and $N=1$ in the Jacobi identity. We finely find that the
most general solution of the eq (20)is the form \beq
D_{M,N}=\frac{d}{12} M(M^{2}-1) \delta_{M+N=0},\eeq where $d$ is
some constant.\\

Hence, by imposing that the generators
$\gamma^{m_{1}}_{i}\gamma^{m_{2}}_{j} \sim e^{\bar{m}}_{(i,j)}$
satisfying the area preserving algebra with non-trivial charge, we
have see that the embedding of the classical Virasoro algebra into
generalized Clifford algebras can be extended to the quantum
Virasoro algebra. Finally, it interesting to
 modified the construction giving the $w_{\infty}$-algebra in terms of the generators
 $\gamma_{i}$ algebra and leading to the central extensions.
\section*{Acknowledgements}
\hspace{.3in} The author is grateful to the Abdus-Salam ICPT,
Trieste, Italy for its hospitality, where this work was done.

\end{document}